\documentclass[aps,prb,twocolumn,showpacs]{revtex4}

\usepackage{graphicx}
\usepackage{amsmath}
\usepackage{amssymb}

\begin{document}

\title{Magnetic field induced reduction of the low-temperature
superfluid density in cuprate superconductors}

\author{Zheyu Huang, Huaisong Zhao, and Shiping Feng$^{*}$}
\affiliation{Department of Physics, Beijing Normal University,
Beijing 100875, China}

\begin{abstract}
The weak magnetic field induced reduction of the low-temperature
superfluid density in cuprate superconductors is studied based on
the kinetic energy driven superconducting mechanism. The
electromagnetic response kernel is evaluated by considering both
couplings of the electron charge and electron magnetic momentum with
a weak magnetic field and employed to calculate the superfluid
density, then the main features of the weak magnetic field induced
reduction of the low-temperature superfluid density are
qualitatively reproduced. The theory also shows that the striking
behavior of the weak magnetic field induced reduction of the
low-temperature superfluid density is intriguingly related to both
depairing due to the Pauli spin polarization and nonlocal response
in the vicinity of the d-wave gap nodes on the Fermi surface to a
weak magnetic field.
\end{abstract}


\maketitle

The superfluid density $\rho_{s}$, being proportional to the density
of the supercarriers, is one of the important characteristic of the
superconducting (SC) condensate \cite{schrieffer83}. It is sensitive
to the low-lying excitation spectrum, and therefore the knowledge of
the superfluid density is essential to understanding the physics of
the underlying mechanism responsible for superconductivity
\cite{schrieffer83,bonn96}. Since cuprate superconductors are doped
Mott insulators with the strong short-range antiferromagnetic
correlation dominating the entire SC phase \cite{damascelli03}, the
magnetic field can be also used to probe the doping and momentum
dependence of the SC gap and spin structure of the Cooper pair
\cite{bonn96}. This is why the first evidence of the d-wave Cooper
pairing state in cuprate superconductors was obtained from the
experimental measurement for the magnetic field penetration depth
$\lambda$ (then the superfluid density $\rho_{\rm s}\equiv
\lambda^{-2}$) \cite{r4}. Experimentally, by virtue of systematic
studies using the muon-spin-rotation measurement technique, some
essential features of the superfluid density in cuprate
superconductors have been established now for all the temperature
$T\leq T_{c}$ throughout the SC dome
\cite{r4,r5,r6,sonier94,r7,sonier99,bidinosti99,r10,k09,k10,serafin10}.
However, there are numerous anomalies, which complicate the physical
properties of the superfluid density. Among these anomalies is the
magnetic field dependence of the superfluid density first observed
on the cuprate superconductor YBa$_{2}$Cu$_{3}$O$_{7-\delta}$
\cite{sonier94,r7,sonier99,bidinosti99}, where a weak magnetic field
can induce an reduction of the superfluid density at the low
temperatures. Later, this weak magnetic field induced reduction of
the low-temperature superfluid density was also found in other
families of cuprate superconductors \cite{r10,k09,k10,serafin10}.

The appearance of the weak magnetic field induced reduction of the
low-temperature superfluid density in cuprate superconductors is the
mostly remarkable effect, however, its full understanding is still a
challenging issue. The earlier work gave the main impetus for a
phenomenological description of the magnetic field dependent
superfluid density in the Meissner state, where it has been argued
that the weak magnetic field induced reduction of the
low-temperature superfluid density arises from nonlinear response of
the d-wave state to a weak magnetic field \cite{yip92}. Later, a
more dominant contribution to the magnetic field dependence of the
single-particle excitations in the superfluid density comes from the
nonlocality of the supercurrent response in the vicinity of the
d-wave gap nodes on the Fermi surface in the d-wave SC state
\cite{amin98}. In our recent work \cite{feng10} based on the kinetic
energy driven SC mechanism \cite{feng0306}, the doping and
temperature dependence of the electromagnetic response in cuprate
superconductors has been discussed by considering the coupling of
the electron charge with a weak magnetic field, where the Meissner
effect is obtained for all the temperature $T\leq T_{c}$ throughout
the SC dome, and then the main features of the doping and
temperature dependence of the local magnetic field profile, the
magnetic field penetration depth, and the superfluid density are
well reproduced. In particular, it is shown that in analogy to the
domelike shape of the doping dependent SC transition temperature,
the maximal superfluid density occurs around the critical doping
$\delta\approx 0.195$, and then decreases in both lower doped and
higher doped regimes. However, the coupling of the electron magnetic
momentum with a weak magnetic field in terms of the Zeeman mechanism
has been dropped in these discussions \cite{feng10}. In this paper,
we study the weak magnetic field induced reduction of the
low-temperature superfluid density in cuprate superconductors by
considering both couplings of the electron charge and electron
magnetic momentum with a weak magnetic field. Following the linear
response theory, we have evaluated the magnetic field dependence of
the response kernel within the kinetic energy driven SC mechanism.
This response kernel is employed to calculate the superfluid
density, then the main features of the weak magnetic field induced
reduction of the low-temperature superfluid density are
qualitatively reproduced. Our results also show that the striking
behavior of the weak magnetic field induced reduction of the
low-temperature superfluid density is intriguingly related to both
depairing due to the Pauli spin polarization and nonlocal response
in the vicinity of the d-wave gap nodes on the Fermi surface to a
weak magnetic field.

We start from the $t$-$J$ model on a square lattice
\cite{anderson87}. However, for discussions of the weak magnetic
field induced reduction of the low-temperature superfluid density in
cuprate superconductors, the $t$-$J$ model can be extended by
including the exponential Peierls factor and Zeeman term as,
\begin{eqnarray}\label{tjham}
H&=&-t\sum_{l\hat{\eta}\sigma}e^{-i{e\over\hbar}{\bf A}(l)\cdot
\hat{\eta}}C^{\dagger}_{l\sigma}C_{l+\hat{\eta}\sigma}\nonumber\\
&+& t'\sum_{l\hat{\eta}'\sigma}e^{-i{e\over\hbar}{\bf A}(l)\cdot
\hat{\eta}'}C^{\dagger}_{l\sigma}C_{l+\hat{\eta}'\sigma}+\mu
\sum_{l\sigma} C^{\dagger}_{l\sigma}C_{l\sigma}\nonumber\\
&+&J\sum_{l\hat{\eta}}{\bf S}_{l}\cdot {\bf S}_{l+\hat{\eta}}
-\varepsilon_{B}\sum_{l\sigma}\sigma C^{\dagger}_{l\sigma}
C_{l\sigma},
\end{eqnarray}
supplemented by an important on-site local constraint $\sum_{\sigma}
C^{\dagger}_{l\sigma}C_{l\sigma}\leq 1$ to remove the double
occupancy, where $\hat{\eta}=\pm\hat{x},\pm\hat{y}$, $\hat{\eta}'
=\pm\hat{x}\pm\hat{y}$, $C^{\dagger}_{l\sigma}$ ($C_{l\sigma}$) is
the electron creation (annihilation) operator, ${\bf S}_{l}=
(S^{x}_{l},S^{y}_{l}, S^{z}_{l})$ are spin operators, and $\mu$ is
the chemical potential. The exponential Peierls factors account for
the coupling of the electron charge to a weak magnetic field in
terms of the vector potential ${\bf A}(l)$, while the Zeeman
magnetic energy $\varepsilon_{B}= g\mu_{B}B$ accounts for the
coupling of the electron magnetic momentum $g\mu_{B}$ with the weak
magnetic field ${\bf B}=\rm{rot}{\bf A}$, with the Lande factor $g$
and Bohr magneton $\mu_{B}$. For a proper description of the
electron single occupancy local constraint, the charge-spin
separation (CSS) fermion-spin theory \cite{feng0304} has been
proposed, where the physics of no double occupancy is taken into
account by representing the electron as a composite object created
by $C_{l\uparrow}= h^{\dagger}_{l\uparrow} S^{-}_{l}$ and
$C_{l\downarrow}=h^{\dagger}_{l\downarrow} S^{+}_{l}$, with the
spinful fermion operator $h_{l\sigma}= e^{-i\Phi_{l\sigma}}h_{l}$
that describes the charge degree of freedom of the electron together
with some effects of spin configuration rearrangements due to the
presence of the doped hole itself (charge carrier), while the spin
operator $S_{l}$ represents the spin degree of freedom of the
electron, then the electron single occupancy local constraint is
satisfied in analytical calculations. In this CSS fermion-spin
representation, the $t$-$J$ model (\ref{tjham}) can be expressed as,
\begin{eqnarray}\label{cssham}
&H&=t\sum_{l\hat{\eta}}e^{-i{e\over\hbar}{\bf A}(l)\cdot\hat{\eta}}
(h^{\dagger}_{l+\hat{\eta}\uparrow}h_{l\uparrow}S^{+}_{l}
S^{-}_{l+\hat{\eta}}+h^{\dagger}_{l+\hat{\eta}\downarrow}
h_{l\downarrow}S^{-}_{l}S^{+}_{l+\hat{\eta}})\nonumber\\
&-&t'\sum_{l\hat{\eta}'} e^{-i{e\over\hbar}{\bf A}(l)\cdot
\hat{\eta}'}(h^{\dagger}_{l+\hat{\eta}'\uparrow}h_{l\uparrow}
S^{+}_{l}S^{-}_{l+\hat{\eta}'}
+h^{\dagger}_{l+\hat{\eta}'\downarrow}h_{l\downarrow}S^{-}_{l}
S^{+}_{l+\hat{\eta}'})\nonumber\\
&-&\mu\sum_{l\sigma} h^{\dagger}_{l\sigma} h_{l\sigma}+J_{{\rm eff}}
\sum_{l\hat{\eta}}{\bf S}_{l}\cdot {\bf S}_{l+\hat{\eta}}-
2\varepsilon_{B}\sum_{l}S^{z}_{l},
\end{eqnarray}
where $J_{{\rm eff}}=(1-\delta)^{2}J$, and $\delta=\langle
h^{\dagger}_{l\sigma}h_{l\sigma}\rangle=\langle h^{\dagger}_{l}
h_{l}\rangle$ is the doping concentration.

For a microscopic description of the SC state of cuprate
superconductors, the kinetic energy driven SC mechanism
\cite{feng0306} has been developed based on the CSS fermion-spin
theory, where the charge carrier-spin interaction from the kinetic
energy term in the $t$-$J$ model (\ref{cssham}) induces a charge
carrier d-wave pairing state by exchanging spin excitations in the
higher power of the doping concentration, then the electron Cooper
pairs originating from the charge carrier pairing state are due to
the charge-spin recombination, and their condensation reveals the
d-wave SC ground-state. In particular, this kinetic energy driven SC
state is the conventional Bardeen-Cooper-Schrieffer (BCS)-like with
the d-wave symmetry, so that all main low energy features of the SC
coherence of the quasiparticle peaks have been quantitatively
reproduced \cite{guo07,lan07}, although the pairing mechanism is
driven by the kinetic energy by exchanging spin excitations. For
discussions of the weak magnetic field induced reduction of the
low-temperature superfluid density, we generalize the analytical
calculation from our previous case \cite{feng10} without considering
the coupling of the electron magnetic momentum with a weak magnetic
field to the case in the presence of the coupling of the electron
magnetic momentum with a weak magnetic field. Following the
discussions in Ref. \onlinecite{feng0306} and Ref.
\onlinecite{guo07}, the mean-field (MF) spin excitation spectrum in
the $t$-$J$ model (\ref{cssham}) in the presence of the coupling of
the electron magnetic momentum with a weak magnetic field can be
evaluated as $\omega^{(B)}_{\bf p}= \sqrt{\omega^{2}_{\bf p}+
(2\varepsilon_{B})^{2}}$, with $\omega_{{\bf p}}$ is the spin
excitation spectrum in the case without the coupling of the electron
magnetic momentum with a weak magnetic field, and has been given in
Ref. \onlinecite{guo07}. Obviously, an additional spin gap
$2\varepsilon_{B}=2g\mu_{B}B$ in the spin excitation spectrum is
induced by the externally applied magnetic field ${\bf B}$. In this
case, the full charge carrier Green's function can be obtained
explicitly in the Nambu representation as,
\begin{eqnarray}
\mathbb{G}({\bf k},i\omega_{n},B)=Z^{(B)}_{\rm hF}{i\omega_{n}
\tau_{0}+\bar{\xi}_{\bf k}\tau_{3}- \bar{\Delta}^{(B)}_{\rm hZ}
({\bf k})\tau_{1}\over (i\omega_{n})^{2}-E^{(B)2}_{{\rm h}{\bf k}}},
\label{holegreenfunction}
\end{eqnarray}
where $\tau_{0}$ is the unit matrix, $\tau_{1}$ and $\tau_{3}$ are
Pauli matrices, the renormalized charge carrier excitation spectrum
$\bar{\xi}_{\bf k}=Z_{\rm hF}\xi_{\bf k}$, with the MF charge
carrier excitation spectrum $\xi_{\bf k}=Zt\chi_{1} \gamma_{\bf
k}-Zt'\chi_{2}\gamma_{\bf k}'- \mu$, the spin correlation functions
$\chi_{1}=\langle S_{i}^{+} S_{i+\hat{\eta}}^{-}\rangle$, $\chi_{2}=
\langle S_{i}^{+} S_{i+\hat{\eta}'}^{-}\rangle$, $\gamma_{\bf k}=
(1/Z) \sum_{\hat{\eta}}e^{i{\bf k}\cdot\hat{\eta}}$, $\gamma_{\bf
k}'= (1/Z)\sum_{\hat{\eta}'}e^{i{\bf k}\cdot\hat{\eta}'}$, $Z$ is
the number of the nearest neighbor or next-nearest neighbor sites,
the renormalized charge carrier d-wave pair gap
$\bar{\Delta}^{(B)}_{\rm hZ} ({\bf k})=Z^{(B)}_{\rm hF}
\bar{\Delta}^{(B)}_{\rm h}({\bf k})$, with the effective charge
carrier d-wave pair gap $\bar{\Delta}^{(B)}_{\rm h}({\bf k})=
\bar{\Delta}^{(B)}_{\rm h}({\rm cos}k_{x}-{\rm cos} k_{y})/2$, and
the charge carrier quasiparticle spectrum $E^{(B)}_{{\rm h}{\bf k}}
=\sqrt{\bar{\xi}^{2}_{{\bf k}}+ |\bar{\Delta}^{(B)}_{\rm hZ}({\bf k}
)|^{2}}$, while the magnetic field dependence of the effective
charge carrier pair gap $\bar{\Delta}^{(B)}_{\rm h}({\bf k})$ and
the quasiparticle coherent weight $Z^{(B)}_{\rm hF}$ satisfy the
following two equations $\bar{\Delta}^{(B)}_{h}({\bf
k})=\Sigma^{(h)}_{2}({\bf k},\omega=0, B)$ and $Z^{(B)-1}_{\rm
hF}(B)=1- \Sigma^{(h)}_{1{\rm o}}({\bf k}, \omega=0,B)\mid_{{\bf
k}=[\pi,0]}$, where $\Sigma^{(h)}_{1}({\bf k}, \omega,B)$ and
$\Sigma^{(h)}_{2}({\bf k},\omega,B)$ are the charge carrier
self-energies obtained from the spin bubble, while
$\Sigma^{(h)}_{1{\rm o}}({\bf k},\omega,B)$ is the antisymmetric
part of $\Sigma^{(h)}_{1}({\bf k},\omega,B)$. Moreover, the forms of
the obtained $\Sigma^{(h)}_{1}({\bf k},\omega,B)$ and
$\Sigma^{(h)}_{2}({\bf k}, \omega,B)$ in the present case are the
same as these given in Ref. \onlinecite{guo07} for the case without
considering the coupling of the electron magnetic momentum with a
weak magnetic field except the spin excitation spectrum
$\omega_{{\bf p}}$ has been replaced by $\omega^{(B)}_{{\bf p}}$,
therefore the effect of a weak magnetic field ${\bf B}$ on the SC
state is considered explicitly through the weak magnetic field ${\bf
B}$ entering in $\Sigma^{(h)}_{1}({\bf k},\omega,B)$ and
$\Sigma^{(h)}_{2}({\bf k}, \omega,B)$. These two equations should be
solved simultaneously with other self-consistent equations
\cite{feng0306,guo07}. However, in the limit of the weak magnetic
field $B\ll B_{c}$ for a given doping concentration, the main role
of the coupling of the electron magnetic momentum with a weak
magnetic field is induced an additional spin gap in the spin
excitation spectrum as we have mentioned above, while the effect on
other spin correlation functions is negligible. In this case, we
will concentrate on the effect of a weak magnetic field on the
charge carrier part since the SC state is dominated by the charge
carrier d-wave pairing state, where the effective charge carrier gap
parameter $\bar{\Delta}^{(B)}_{\rm h}$, the quasiparticle coherent
weight $Z^{(B)}_{\rm hF}$, the chemical potential, and other charge
carrier particle-hole parameters must be solved self-consistently.

In the presence of the coupling of the electron magnetic momentum
with a weak magnetic field, the magnetic field dependence of the
response current density $J_{\mu}$ and the vector potential
$A_{\nu}$ are related by a nonlocal kernel of the response function
$K_{\mu\nu}$ as \cite{feng10},
\begin{equation}\label{linres}
J_{\mu}({\bf q},\omega,B)=-\sum\limits_{\nu=1}^{3} K_{\mu\nu}({\bf
q},\omega,B)A_{\nu}({\bf q},\omega),
\end{equation}
with the Greek indices label the axes of the Cartesian coordinate
system. This magnetic field dependence of the response kernel
(\ref{linres}) can be separated into two parts as $K_{\mu\nu}({\bf
q},\omega,B)=K^{({\rm d})}_{\mu\nu}({\bf q},\omega,B)+K^{({\rm p})
}_{\mu\nu} ({\bf q}, \omega,B)$, where $K^{({\rm d})}_{\mu\nu}({\bf
q},\omega,B)$ and $K^{({\rm p} )}_{\mu\nu}({\bf q},\omega,B)$ are
the corresponding diamagnetic and paramagnetic parts, respectively.
In the CSS fermion-spin representation \cite{feng0304}, the vector
potential ${\bf A}$ has been coupled to the electron charge, which
are now represented by $C_{l\uparrow}= h^{\dagger}_{l\uparrow}
S^{-}_{l}$ and $C_{l\downarrow}=h^{\dagger}_{l\downarrow}S^{+}_{l}$.
In this case, the electron polarization operator is expressed as
${\bf P}=-e\sum \limits_{i\sigma}{\bf R}_{i} C^{\dagger}_{i\sigma}
C_{i\sigma}=e\sum\limits_{i\sigma}{\bf R}_{i} h^{\dagger}_{i}
h_{i}$, then the corresponding electron current operator is obtained
by evaluating the time-derivative of this polarization operator as
${\bf j}={\bf j}^{(d)}+{\bf j}^{(p)}$, with ${\bf j}^{(d)}$ and
${\bf j}^{(p)}$ are the corresponding diamagnetic (d) and
paramagnetic (p) components of the electron current operator.
Following our previous discussions \cite{feng10}, these diamagnetic
and paramagnetic parts of the magnetic field dependence of the
response kernel $K^{({\rm d})}_{\mu\nu}({\bf q},\omega,B)$ and
$K^{({\rm p} )}_{\mu\nu}({\bf q},\omega,B)$ can be obtained in the
the static limit as,
\begin{subequations}\label{allkernel}
\begin{eqnarray}
K_{\mu\nu}^{(\rm{d})}({\bf q},0,B)&=&-{4e^{2}\over\hbar^{2}}
(\chi_{1}\phi_{1}t-2\chi_{2}\phi_{2}t')\delta_{\mu\nu}={1\over
\lambda^{2}_{L}}\delta_{\mu\nu},\label{diakernel} \\
K_{\mu\nu}^{(\rm{p})}({\bf q},0,B)&=&{1\over N}\sum\limits_{{\bf k}}
{\bf \gamma}_{\mu}({\bf k}+{\bf q},{\bf k}) {\bf \gamma}^{*}_{\nu}
({\bf k}+{\bf q},{\bf k})[L^{(B)}_{1}({\bf k},{\bf q})\nonumber\\
&+&L^{(B)}_{2}({\bf k},{\bf q})]=K_{\mu\mu}^{(\rm{p})}({\bf q},0,B)
\delta_{\mu\nu}, ~~~~~~~~~~\label{parakernel}
\end{eqnarray}
\end{subequations}
where the charge carrier particle-hole parameters $\phi_{1}=\langle
h^{\dagger}_{i\sigma} h_{i+\hat{\eta}\sigma}\rangle$ and $\phi_{2}=
\langle h^{\dagger}_{i\sigma}h_{i+\hat{\eta}'\sigma} \rangle$,
$\lambda^{-2}_{L}=-4e^{2}(\chi_{1}\phi_{1}t- 2\chi_{2} \phi_{2}t')
/\hbar^{2}$ is the London penetration depth, and now is doping,
temperature, and magnetic field dependent, the bare current vertex
${\bf \gamma}_\mu({\bf k}+{\bf q},{\bf k})$ has been given in Ref.
\onlinecite{feng10}, while the functions $L^{(B)}_{1}({\bf k},{\bf
q})$ and $L^{(B)}_{2}({\bf k},{\bf q})$ are obtained as,
\begin{widetext}
\begin{subequations}\label{lfunctions}
\begin{eqnarray}
L^{(B)}_{1}({\bf k},{\bf q})&=&Z^{(B)2}_{\rm hF}\left(1+
{\bar{\xi}_{\bf k}\bar{\xi}_{{\bf k}+{\bf q}}+
\bar{\Delta}^{(B)}_{\rm hZ}({\bf k}) \bar{\Delta}^{(B)}_{\rm hZ}
({\bf k}+{\bf q})\over E^{(B)}_{{\rm h}{\bf k}}E^{(B)}_{{\rm h}{{\bf
k}+{\bf q}}}}\right){n_{F}(E^{(B)}_{{\rm h} {\bf
k}})-n_{F}(E^{(B)}_{{\rm h}{{\bf k}+{\bf q}}})\over E^{(B)}_{{\rm
h}{\bf k}}-E^{(B)}_{{\rm h}{{\bf k}+{\bf q}}}},
~~~~~~~~\\
L^{(B)}_{2}({\bf k},{\bf q})&=&Z^{(B)2}_{\rm hF}\left(1-
{\bar{\xi}_{\bf k} \bar{\xi}_{{\bf k}+{\bf q}}+
\bar{\Delta}^{(B)}_{\rm hZ}({\bf k}) \bar{\Delta}^{(B)}_{\rm hZ}
({\bf k}+{\bf q})\over E^{(B)}_{{\rm h}{\bf k}}E^{(B)}_{{\rm h}{{\bf
k}+{\bf q}}}}\right){n_{F}(E^{(B)}_{{\rm h} {\bf k}})+
n_{F}(E^{(B)}_{{\rm h}{{\bf k}+{\bf q}}})-1\over E^{(B)}_{{\rm h}
{\bf k}}+E^{(B)}_{{\rm h}{{\bf k}+{\bf q}}}}. ~~~~~~~~
\end{eqnarray}
\end{subequations}
\end{widetext}
It is easy to show \cite{feng10} that in the long wavelength limit,
i.e., $|{\bf q}|\to 0$, $K_{yy}^{({\rm p})}({\bf q}\to 0,0,B)=0$ at
the temperature $T=0$. In this case, the long wavelength
electromagnetic response is determined by the diamagnetic part of
the kernel only. On the other hand, at the SC transition temperature
$T=T_{c}$, $K_{yy}^{({\rm p} )}({\bf q}\to 0,0,B)=-(1/
\lambda^{2}_{L})$, which exactly cancels the diamagnetic part of the
response kernel (\ref{diakernel}), and then the Meissner effect in
the presence of the coupling of the electron magnetic momentum with
a weak magnetic field is obtained for all $T\leq T_{c}$ throughout
the SC dome.

However, the result we have obtained the response kernel in Eqs.
(\ref{diakernel}) and (\ref{parakernel}) can not be used for a
direct comparison with the corresponding experimental data of
cuprate superconductors because the response kernel derived within
the linear response theory describes the response of an
\emph{infinite} system, whereas in the problem of the penetration of
the field and the system has a surface, i.e., it occupies a
half-space $x>0$. In such problems, it is necessary to impose
boundary conditions for charge carriers. This can be done within the
simplest specular reflection model \cite{landau80} with a
two-dimensional (2D) geometry of the SC plane. Taking into account
the 2D geometry of cuprate superconductors within the specular
reflection model \cite{landau80}, we \cite{feng10} can obtain the
magnetic field penetration depth as,
\begin{eqnarray}\label{lambda}
\lambda(T,B)&=&{1\over B}\int\limits_{0}^{\infty}h_{z}(x,B)\,{\rm
d}x\nonumber\\
&=& {2\over\pi}\int\limits_{0}^{\infty}{{\rm d}q_{x}\over\mu_{0}
K_{yy}(q_{x},0,0,B)+q_{x}^{2}},
\end{eqnarray}
which therefore reflects the measurably electromagnetic response in
cuprate superconductors.

\begin{figure}[h!]
\includegraphics[scale=0.45]{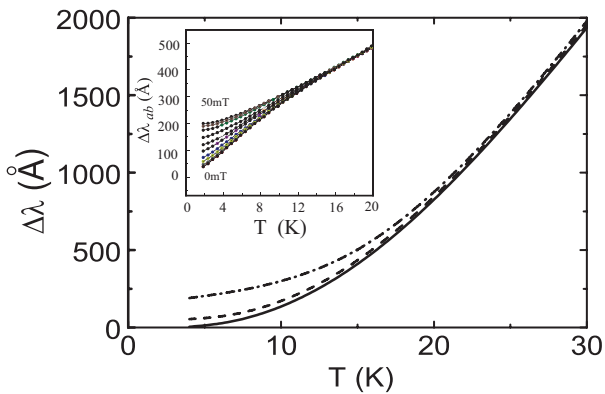}
\caption{The magnetic field penetration depth $\Delta\lambda$ as a
function of temperature at the doping concentration $\delta=0.09$
for the magnetic field $B=0$ T (solid line), $B=0.5$ T (dashed
line), and $B=1.0$ T (dash-dotted line) with parameters $t/J=2.5$,
$t'/t=0.3$, and $J=1000$K. Inset: the corresponding experimental
data for YBa$_{2}$Cu$_{4}$O$_{8}$ taken from Ref.
\onlinecite{serafin10}. \label{lambdafig}}
\end{figure}

Now we are ready to discuss the magnetic field dependence of the
Meissner effect in cuprate superconductors. In cuprate
superconductors, although the values of $J$, $t$, and $t'$ are
believed to vary somewhat from compound to compound
\cite{damascelli03}, however, as a qualitative discussion as in our
recent work \cite{feng10}, the commonly used parameters
\cite{hybertson90,pavarini01} in this paper are chosen as $t/J=2.5$,
$t'/t=0.3$, and $J=1000$K. Furthermore, a characteristic length
scale $a_{0}= \sqrt{\hbar^{2} a/\mu_{0}e^{2}J}$ is introduced. Using
the lattice parameter $a\approx 0.383$nm for YBa$_2$Cu$_3$O$_{7-y}$,
this characteristic length is obtain as $a_{0}\approx 97.8$nm. In
this case, the magnetic field penetration depth $\Delta\lambda(T,B)
=\lambda(T,B) -\lambda(0,0)$ as a function of temperature at the
doping concentration $\delta=0.09$ for the magnetic field $B=0$ T
(solid line), $B=0.5$ T (dashed line), and $B=1.0$ T (dash-dotted
line) is plotted in Fig. \ref{lambdafig} in comparison with the
corresponding experimental results \cite{serafin10} of
YBa$_{2}$Cu$_{4}$O$_{8}$ (inset). The similar magnetic field
dependence of the magnetic field penetration depth has been also
observed experimentally on YBa$_{2}$Cu$_{3}$O$_{6.95}$
\cite{sonier94}. However, YBa$_{2}$Cu$_{3}$O$_{7-\delta}$ contains a
single CuO chain per unit cell whose oxygen content can be varied to
tune the doping level on the CuO$_{2}$ plane \cite{sonier94}, while
its close relative YBa$_{2}$Cu$_{4}$O$_{8}$ has a double chain layer
that is stoichiometric and a planar state that is underdoped
\cite{serafin10}. Although this difference of the electronic
structure of the quasi-2D plane leads to some subtly different
behaviors between YBa$_{2}$Cu$_{3}$O$_{7-\delta}$ and
YBa$_{2}$Cu$_{4}$O$_{8}$, where at a weak magnetic field, the
experimental curves of the temperature dependent magnetic field
penetration depth for YBa$_{2}$Cu$_{3}$O$_{6.95}$ show curvature in
low temperature \cite{sonier94} as in the case of zero magnetic
field, while the corresponding experimental curves of the
temperature dependent magnetic field penetration depth for
YBa$_{2}$Cu$_{4}$O$_{8}$ are linear \cite{serafin10}, the
qualitative properties in both YBa$_{2}$Cu$_{3}$O$_{7-\delta}$ and
YBa$_{2}$Cu$_{4}$O$_{8}$ are consistent each other
\cite{sonier94,serafin10}. In this paper, we mainly focus on the
qualitative properties of the magnetic field dependent Meissner
effect in cuprate superconductors based on the simple $t$-$J$ model
(\ref{tjham}). Within the kinetic energy driven SC mechanism, the SC
transition temperature $T_{c}=52$K at $\delta=0.09$ for zero
magnetic field. As we \cite{feng10} have shown that at the SC
transition temperature $T=T_{c}$, the kernel of the response
function $K_{\mu\nu}({\bf q} \to 0,0,0)|_{T=T_{c}}=0$. In this case,
we obtain the magnetic field penetration depth from Eq.
(\ref{lambda}) as $\lambda(T_{c},0)=\infty$, which reflects that in
the normal state, the external magnetic field can penetrate through
the main body of the system, therefore there is no the Meissner
effect in the normal state. Furthermore, our present result in Fig.
\ref{lambdafig} shows clearly that the characteristic feature of the
temperature dependent $\lambda(T,B)$ is essentially independent on a
weak magnetic field, in particular, $\lambda(T,B)$ shows a crossover
from the linear temperature dependence at higher temperatures to a
nonlinear one in the low temperatures as in the case of zero
magnetic field \cite{feng10}, in qualitative agreement with the
corresponding experimental data \cite{sonier94} of
YBa$_{2}$Cu$_{3}$O$_{6.95}$. Moreover, the magnitude of
$\lambda(T,B)$ at the low temperatures is dependent on a weak
magnetic field, and then it is independent on a weak magnetic field
at higher temperatures. However, there is a substantial difference
between theory and experiment, namely, the value of the magnetic
field dependent penetration depth $\Delta\lambda(T,B)$ at the low
temperatures calculated theoretically is smaller than the
corresponding value measured in the experiment
\cite{serafin10,sonier94}. However, upon a closer examination one
can see immediately that the main difference is due to fact that the
calculated value of $\Delta\lambda(T,B)$ increases slowly with
magnetic field at the low temperatures. The simple $t$-$J$ model can
not be regarded as a complete model for the quantitative comparison
with cuprate superconductors, however, as for a qualitative
discussion in this paper, the overall shape seen in the theoretical
result is qualitatively consistent with that observed in the
experiment \cite{sonier94,serafin10}.

\begin{figure}[h!]
\includegraphics[scale=0.5]{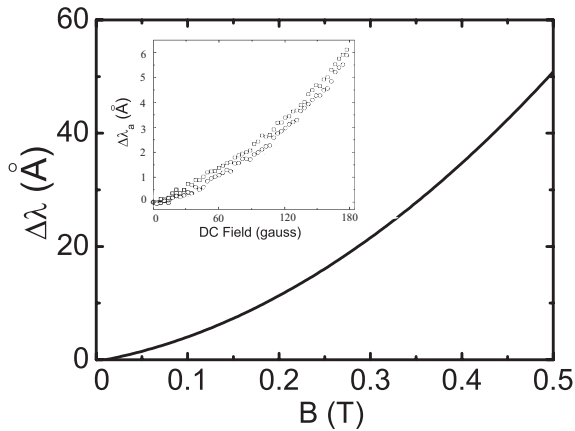}
\caption{The magnetic field penetration depth as a function of
magnetic field at the doping concentration $\delta=0.09$ for
temperature $T=4$K with parameters $t/J=2.5$, $t'/t=0.3$, and
$J=1000$K. Inset: the corresponding experimental data for
YBa$_{2}$Cu$_{3}$O$_{6.95}$ at temperature $T=4.2$K (circles) and
$T=7$K (squares) taken from Ref. \onlinecite{bidinosti99}.
\label{lambdafig1}}
\end{figure}

To show this magnetic field dependence of $\lambda(T,B)$ at the low
temperatures clearly, we have made a series of calculations for
$\lambda(T,B)$ at differently weak magnetic fields, and the result
of $\Delta\lambda(T,B)$ as a function of magnetic field at
$\delta=0.09$ with $T=4$K is plotted in Fig. \ref{lambdafig1} in
comparison with the corresponding experimental results
\cite{bidinosti99} of YBa$_2$Cu$_3$O$_{6.95}$ at temperature
$T=4.2$K (circles) and $T=7$K (squares) (inset). Obviously,
$\lambda(T,B)$ is a nonlinear function of magnetic field. Moreover,
a rapid increase of $\lambda(T,B)$ with a weak magnetic field at the
low temperatures observed from cuprate superconductors
\cite{bidinosti99,sonier99} is qualitatively reproduced. It should
be emphasized that the present result for the d-wave SC state in
cuprate superconductors is much different from that in the
conventional superconductors, where the magnetic field dependence is
typically weak at the low temperatures because the isotropic energy
gap exponentially cut off the quasiparticle excitations.

\begin{figure}[h!]
\includegraphics[scale=0.5]{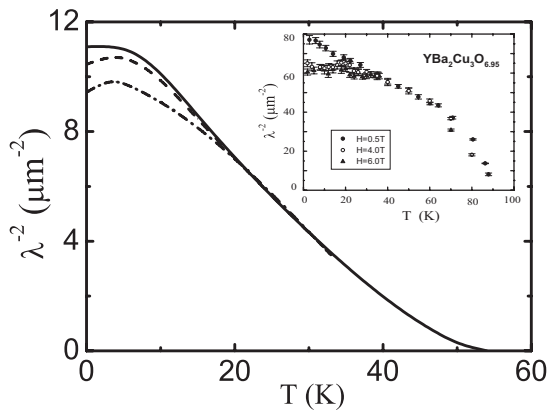}
\caption{The superfluid density as a function of temperature at the
doping concentration $\delta=0.09$ for the magnetic field $B=0$ T
(solid line), $B=0.5$ T (dashed line), and $B=1.0$ T (dash-dotted
line) with parameters $t/J=2.5$, $t'/t=0.3$, and $J=1000$K. Insets:
the corresponding experimental results for YBa$_2$Cu$_3$O$_{6.95}$
taken from Ref. \onlinecite{sonier99}. \label{rhofig}}
\end{figure}

Next we discuss the weak magnetic field induced reduction of the
low-temperature superfluid density $\rho_{\rm s}(T,B)$. This
superfluid density is closely related to the magnetic field
penetration depth as $\rho_{\rm s}(T,B)\equiv \lambda^{-2}(T,B)$. In
this case, we have made a series of calculations for $\rho_{\rm
s}(T,B)$ at differently weak magnetic fields, and the result of
$\rho_{\rm s}(T,B)$ as a function of temperature at $\delta=0.09$
for $B=0$ T (solid line), $B=0.5$ T (dashed line), and $B=1.0$ T
(dash-dotted line) is plotted in Fig. \ref{rhofig} in comparison
with the corresponding experimental data for YBa$_2$Cu$_3$O$_{6.95}$
taken from Ref. \onlinecite{sonier99} (inset). In correspondence
with the result of $\lambda(T,B)$ shown in Fig. \ref{lambdafig}, the
characteristic feature of the temperature dependent superfluid
density is also independent on a weak magnetic field, where as in
the case of zero magnetic field, $\rho_{\rm s}(T,B)$ shows a linear
temperature dependence at higher temperatures, and then it crosses
over to a nonlinear temperature behavior at the low temperatures.
However, most importantly, the magnitude of $\rho_{\rm s}(T,B)$ at
the low temperatures decreases with increasing magnetic field, and
then it turns to be independent on a weak magnetic field at higher
temperatures, in qualitative agreement with experimental data of
cuprate superconductors \cite{r7,sonier99,r10,k09,k10}. The present
result also indicates that the nature of the quasiparticle
excitations at the low temperatures is strongly influenced by a weak
magnetic field. This weak magnetic field induced reduction of the
low-temperature superfluid density in cuprate superconductors
contrasts with the that observed from the conventional
superconductors \cite{khasanov06}, where the curves of the
temperature dependent superfluid density for differently weak
magnetic fields were found to collapse onto a single curve since the
conventional superconductors are fully gaped.

The weak magnetic field induced reduction of the low-temperature
superfluid density in cuprate superconductors arises from both
depairing due to the Pauli spin polarization and nonlocal response
in the vicinity of the d-wave gap nodes on the Fermi surface to a
weak magnetic field. In the framework of the kinetic energy driven
SC mechanism, the d-wave SC state is mediated by the interaction of
electrons and spin excitations \cite{feng0306}, where the depairing
can occur due to the Pauli spin polarization in the presence of an
external magnetic field. This follows a fact that an applied
magnetic field aligns the spins of the unpaired electrons, i.e.,
there is a tendency to induce the magnetic order, then the kinetic
energy driven d-wave Cooper pairs can not take advantage of the
lower energy offered by a spin-polarized state
\cite{feng0306,vorontsov10}. On the other hand, the characteristic
feature of the d-wave superconductors is the existence of four nodes
on the Fermi surface, where the energy gap vanishes
$\bar{\Delta}_{\rm h}({\bf k})|_{\rm at~nodes}= \bar{\Delta}_{\rm
h}({\rm cos}k_{x}-{\rm cos}k_{y})/2 |_{\rm at~ nodes}=0$. These two
special features in the kinetic energy driven d-wave SC state
indicate that even small thermal energy or externally small magnetic
energy can excite excitations, then the superfluid density decreases
with increasing temperature or increasing magnetic field, reflecting
that the weak magnetic field induced reduction of the
low-temperature superfluid density is a natural consequence of the
kinetic energy driven d-wave SC state. Since the quasiparticles
selectively populate the nodal region at the low temperatures, then
the most physical properties in the SC state are controlled by the
quasiparticle excitations around the nodes. In this case, the
Ginzburg--Landau ratio around the nodal region is no longer large
enough for the system to belong to the class of type-II
superconductors, and the condition of the local limit is not
fulfilled \cite{amin98,kosztin97,feng10}. On contrary, the system
falls into the extreme nonlocal limit, then the nonlinear
characteristic in the temperature dependence of the superfluid
density (then the magnetic field penetration depth) can be observed
experimentally in cuprate superconductors at the low temperatures.
However, when a weak magnetic field is applied to the system {\it
even at zero temperature}, the quasiparticles around the nodal
region become excited out of the condensate, and at the same time
the electron attractive interaction for the Cooper pairs by
exchanging spin excitations is weaken \cite{feng0306}, both these
effects lead to a decreases in the superfluid density. With
increasing temperatures, the externally small magnetic energy due to
the presence of a weak magnetic field is comparable with the small
thermal energy at the low temperatures, therefore both small thermal
energy and weak magnetic field induce an reduction of the superfluid
density. However, at higher temperatures, this externally small
magnetic energy is much smaller than the thermal energy, then the
major contribution to a decrease of the superfluid density comes
from the thermal energy. This is why a weak magnetic field only
reduces an reduction of the superfluid density only at the low
temperatures.

As we have mentioned in Eq. (\ref{tjham}), the coupling of the
electron charge to a weak magnetic field is in terms of the vector
potential ${\bf A}(l)$, while the coupling of the electron magnetic
momentum $g\mu_{B}$ with the weak magnetic field ${\bf B}=\rm{rot}
{\bf A}$ is in terms of the Zeeman mechanism. During the above
discussions, the electromagnetic response kernel (\ref{allkernel})
is calculated with the bare current vertex \cite{feng10}, where the
important point is that for a weak magnetic field which orientation
is the same at all spatial points the spin polarization axis along
the weak magnetic field is chosen accordingly. As a consequence, we
do not take into account longitudinal excitations properly
\cite{schrieffer83}, and the obtained results are valid only in the
gauge, where the vector potential is purely transverse, e.g. in the
Coulomb gauge. If we want to keep the theory gauge invariant, it is
crucial to approximate the electromagnetic response kernel in a way
maintaining local charge conservation \cite{schrieffer83}. For the
case without considering the coupling of the electron magnetic
momentum with a weak magnetic field, we have shown that although the
electromagnetic response kernel is not manifestly gauge invariant
within the bare current vertex \cite{feng10}, the gauge invariance
is kept within the dressed current vertex \cite{krzyzosiak10}.
However, for the present case with considering both couplings of the
electron charge and electron magnetic momentum with a weak magnetic
field, the gauge invariance should be kept within the dressed
current vertex together with accordingly rotated spin polarization
axis along the weak magnetic field. These and the related issues are
under investigation now.

In conclusion, we have discussed the magnetic field induced
reduction of the low-temperature superfluid density in cuprate
superconductors based on the kinetic energy driven SC mechanism by
considering both couplings of the electron charge and electron
magnetic momentum with a weak magnetic field. Our results show that
although the characteristic feature of the temperature dependent
superfluid density is found to be independent on a weak magnetic
field, this weak magnetic field induces an reduction of the
low-temperature superfluid density in the Meissner state. Our
results also show that the striking behavior of the weak magnetic
field induced reduction of the low-temperature superfluid density
can be attributed to both depairing due to the Pauli spin
polarization and nonlocal response in the vicinity of the d-wave gap
nodes on the Fermi surface to a weak magnetic field.

\acknowledgments

This work was supported by the National Natural Science Foundation
of China under Grant No. 11074023, and the funds from the Ministry
of Science and Technology of China under Grant No. 2011CB921700.


\begin{thebibliography}{00}

\bibitem [*] {add} To whom correspondence should be addressed,
E-mail: spfeng@bnu.edu.cn

\bibitem{schrieffer83} See, e.g., J. R. Schrieffer, \emph{Theory of
Superconductivity} (Addison-Wesley, San Francisco, 1964).

\bibitem {bonn96} See, e.g., B. A. Bonn and W. N. Hardy,
in \emph{Physical Properties of High Temperature Superconductors} V,
edited by D. M. Ginsberg (World Scientific, Singapore, 1996).

\bibitem{damascelli03} See, e.g., A. Damascelli, Z.
Hussain, and Z.-X. Shen, Rev. Mod. Phys. {\bf 75}, 473 (2003).

\bibitem{r4} W. N. Hardy, D. A. Bonn, D. C. Morgan, Ruixing Liang,
and Kuan Zhang, Phys. Rev. Lett. {\bf{70}}, 3999 (1993).

\bibitem{r5}C. Bernhard, J. L. Tallon, Th. Blasius, A. Golnik,
and Ch. Niedermeyer, Phys. Rev. Lett. {\bf{86}}, 1614 (2001).

\bibitem{r6} D. M. Broun, W. A. Huttema, P. J. Turner, S.
\"Ozcan, B. Morgan, Ruixing Liang, W. N. Hardy, and D. A. Bonn,
Phys. Rev. Lett. {\bf{99}}, 237003 (2007).

\bibitem{sonier94} J. E. Sonier, R. F. Kiefl, J. H. Brewer, D. A.
Bonn, J. F. Carolan, K. H. Chow, P. Dosanjh, W. N. Hardy, Ruixing
Liang, W. A. MacFarlane, P. Mendels, G. D. Morris, T. M. Riseman,
and J. W. Schneider, Phys. Rev. Lett. {\bf 72}, 744 (1994).

\bibitem{r7} J. E. Sonier, R. F. Kiefl, J. H. Brewer, D. A. Bonn, S.
R. Dunsiger, W. N. Hardy, Ruixing Liang, W. A. MacFarlane, T. M.
Riseman, D. R. Noakes, and C. E. Stronach, Phys. Rev. B {\bf 55},
11789 (1997).

\bibitem{sonier99} J. E. Sonier, J. H. Brewer, R. F. Kiefl, G. D.
Morris, R. Miller, D. A. Bonn, J. Chakhalian, R. H. Heffner, W. N.
Hardy, and R. Liang, Phys. Rev. Lett. {\bf 83}, 4156 (1999).

\bibitem{bidinosti99} C. P. Bidinosti, W. N. Hardy, D. A. Bonn, and
Ruixing Liang, Phys. Rev. Lett. {\bf 83}, 3277 (1999).

\bibitem{r10} R. Khasanov, A. Shengelaya, A. Maisuradze, F. La
Mattina, A. Bussmann-Holder, H. Keller, and K. A. M\"uller, Phys.
Rev. Lett. {\bf 98}, 057007 (2007).

\bibitem{k09} R. Khasanov, Takeshi Kondo, S. Str\"assle, D. O.
G. Heron, A. Kaminski, H. Keller, S. L. Lee, and Tsunehiro Takeuchi,
Phys. Rev. B {\bf 79}, 180507 (2009).

\bibitem{k10} R. Khasanov, Takeshi Kondo, M. Bendele, Yoichiro
Hamaya, A. Kaminski, S. L. Lee, S. J. Ray, Tsunehiro Takeuchi, Phys.
Rev. B {\bf 82}, 020511 (2010).

\bibitem{serafin10} A. Serafin, J.D. Fletcher, S. Adachi, N.E. Hussey,
and A. Carrington, Phys. Rev. B {\bf 82}, 140506 (2010).

\bibitem{yip92} S. K. Yip and J. Sauls, Phys. Rev. Lett. {\bf 69}, 2264
(1992).

\bibitem{amin98} M. H. S. Amin, Ian Affleck, and M. Franz, Phys. Rev.
B {\bf 58}, 5848 (1998).

\bibitem{feng10} Shiping Feng, Zheyu Huang, and Huaisong Zhao, Physica
C {\bf 470}, 1968 (2010).

\bibitem{feng0306} Shiping Feng, Phys. Rev. B {\bf 68}, 184501 (2003);
Shiping Feng, Tianxing Ma, and Huaiming Guo, Physica C {\bf 436}, 14
(2006).

\bibitem{anderson87} P. W. Anderson, Science {\bf 235}, 1196 (1987).

\bibitem{feng0304} Shiping Feng, Jihong Qin, and Tianxing Ma, J.
Phys.: Condens. Matter {\bf 16},343 (2004).

\bibitem{guo07} Huaiming Guo and Shiping Feng, Phys. Lett. A
{\bf 361}, 382 (2007); Weifang Wang, Zhi Wang, Jingge Zhang, and
Shiping Feng, Phys. Lett. A {\bf 374}, 632 (2010).

\bibitem{lan07} Yu Lan, Jihong Qin, and Shiping Feng, Phys. Rev. B {\bf
76}, 014533 (2007); Zhi Wang and Shiping Feng, Phys. Rev. B {\bf
80}, 064510 (2009).

\bibitem{landau80} See, e.g., A. A. Abrikosov, \emph{Fundamentals of
the Theory of Metals} (Elsevier Science Publishers B. V., 1988).

\bibitem{hybertson90} M.S. Hybertson, E. Stechel, M. Schuter, and
D. Jennison, Phys. Rev. B {\bf 41}, 11068 (1990).

\bibitem{pavarini01} E. Pavarini, I. Dasgupta, T. Saha-Dasgupta, O.
Jepsen, and O. K. Andersen, Phys. Rev. Lett. {\bf 87}, 47003 (2001).

\bibitem{khasanov06} R. Khasanov, I. L. Landau, C. Baines, F. La
Mattina, A. Maisuradze, K. Togano, and H. Keller, Phys. Rev. B {\bf
73}, 214528 (2006).

\bibitem{vorontsov10} A. B. Vorontsov and I. Vekhter, Phys. Rev. B
{\bf 81}, 094527 (2010).

\bibitem{kosztin97}I. Kosztin and A. J. Legget, Phys. Rev. Lett.
{\bf{79}}, 135 (1997).

\bibitem{krzyzosiak10} Mateusz Krzyzosiak, Zheyu Huang, Shiping
Feng, and Ryszard Gonczarek, Physica C {\bf 470}, 407 (2010).

\end{thebibliography}
\end{document}